%% file: main.tex
 \documentclass[conference,10pt]{IEEEtran}
\IEEEoverridecommandlockouts
\usepackage{cite}
\usepackage{amsmath,amssymb,amsfonts}
\usepackage{graphicx}
\usepackage{textcomp}
\usepackage{xcolor}
\usepackage{url}
\usepackage{xurl}
\usepackage[acronym,nomain,nonumberlist,nopostdot,nogroupskip]{glossaries}
\usepackage[nice]{nicefrac}
\usepackage{booktabs}
\usepackage{algorithm}
\usepackage{algorithmic}
\usepackage{subcaption}
\usepackage{fancybox}

\makeatletter
\newcommand\fs@spaceruled{\def\@fs@cfont{\bfseries}\let\@fs@capt\floatc@ruled
  \def\@fs@pre{\vspace{6pt}\hrule height.8pt depth0pt \kern2pt}%
  \def\@fs@post{\kern2pt\hrule\relax}%
  \def\@fs@mid{\kern2pt\hrule\kern2pt}%
  \let\@fs@iftopcapt\iftrue}
\makeatother

\floatstyle{ruled}
\newfloat{spacedalgorithm}{tbp}{loa}
\floatname{spacedalgorithm}{Algorithm}

\begin{document}

\title{Towards Neuromorphic Processing for Next-Generation MU-MIMO Detection}


\author{\IEEEauthorblockN{G. N. Katsaros, J. C. De Luna Ducoing, and Konstantinos Nikitopoulos\\}
\IEEEauthorblockA{5GIC \& 6GIC, Institute for Communication Systems (ICS), University of Surrey, Guildford, UK}}


\maketitle

\begin{abstract}

Upcoming physical layer (PHY) processing solutions, leveraging multiple-input multiple-output (MIMO) advances, are expected to support broad transmission bandwidths and the concurrent transmission of multiple information streams. However, the inherent computational complexities of conventional MIMO PHY algorithms pose significant practical challenges, not only in meeting the strict real-time processing latency requirements but also in maintaining practical computational power consumption budgets.
Novel computing paradigms, such as neuromorphic computing, promise substantial gains in computational power efficiency. However, it is unknown whether it is feasible or efficient to realize practical PHY algorithms on such platforms. In this work, we evaluate for the first time the potential of neuromorphic computing principles for multi-user (MU)-MIMO detection. 
In particular, we developed the first spiking-based MU-MIMO simulator that meets practical error-rate targets, suggesting power gains of at least one order of magnitude when realized on actual neuromorphic hardware, compared to conventional processing platforms. Finally, we discuss the challenges and future research directions that could unlock practical neuromorphic-based MU-MIMO systems and revolutionize PHY power efficiency.



\end{abstract}


\input{0_body}

\bibliographystyle{IEEEtran}
\bibliography{neuromorphic}

\end{document}

%% file: 0_body.tex
\input{abbreviations}

\section{Introduction}\label{s_intro}

\input{1_introduction}

\section{System Model and MIMO Detection Overview}\label{s_sys_mod}
    \input{2_system_model}

\section{QUBO formulation of the ML MIMO Detection Problem}\label{s_adop_meth}
    \input{3_adopted_methodology}

\section{Spiking neurons for MU-MIMO detection}\label{sec_snn_mumimo}
    \input{4_mapping_to_snn}

\section{Results and Discussion}\label{s_results}

\input{5_results_and_discussion}

\section{Challenges and Future directions}\label{s_challenges}
    \input{6_challenges_future_work}

%% file: abbreviations.tex
\newacronym{3gpp}{3GPP}{3rd Generation Partnership Project}
\newacronym{5gnr}{5G-NR}{5G New Radio}
\newacronym{ai}{AI}{artificial intelligence}
\newacronym{amp}{AMP}{approximate message passing}
\newacronym{au}{AU}{allocation unit}
\newacronym{ber}{BER}{bit error rate}
\newacronym{bs}{BS}{base station}
\newacronym{bqm}{BQM}{binary quadratic model}
\newacronym{cdl}{CDL}{clustered delay line}
\newacronym{cfslp}{CF-SLP}{closed-form suboptimal SLP}
\newacronym{dac}{DAC}{digital-to-analog converter}
\newacronym{deepl}{DL}{deep learning}
\newacronym{dmrs}{DM-RS}{demodulation reference signals}
\newacronym{dnn}{DNN}{deep neural network}
\newacronym{csi}{CSI}{channel state information}
\newacronym{fec}{FEC}{forward error correction}
\newacronym{fr1}{FR1}{sub-6 GHz frequency range}
\newacronym{gp}{GP}{Gyre Precoding}
\newacronym{gpu}{GPU}{graphics processing unit}
\newacronym{gzf}{GZF}{greedy zero-forcing}
\newacronym{harq}{HARQ}{hybrid automatic repeat request}
\newacronym{ice}{ICE}{integrated control errors}
\newacronym{ils}{ILS}{integer least-squares}
\newacronym{las}{LAS}{likelihood ascent search}
\newacronym{ldpc}{LDPC}{low-density parity-check}
\newacronym{lif}{LIF}{leaky integrate-and-fire}
\newacronym{llr}{LLR}{log-likelihood ratio}
\newacronym{los}{LOS}{line-of-sight}
\newacronym{lte}{LTE}{Long-Term Evolution}
\newacronym{mcs}{MCS}{modulation and coding scheme}
\newacronym{mf}{MF}{matched filter}
\newacronym{mimo}{MIMO}{multiple-input, multiple-output}
\newacronym{mmimo}{mMIMO}{massive MIMO}
\newacronym{ml}{ML}{maximum likelihood}
\newacronym{mpnl}{MPNL}{Massively Parallel Non-Linear processing}
\newacronym{mmse}{MMSE}{minimum mean square error}
\newacronym{mrc}{MRC}{maximum-ratio combining}
\newacronym{mumimo}{MU-MIMO}{multi-user, multiple-input, multiple-output}
\newacronym{mu}{MU}{multi-user}
\newacronym{nl}{NL}{non-linear}
\newacronym{nlos}{NLOS}{non-line-of-sight}
\newacronym{noma}{NOMA}{non-orthogonal multiple access}
\newacronym{ofdm}{OFDM}{orthogonal frequency-division multiplexing}
\newacronym{opus}{OPUS}{orthogonality probing based user selection}
\newacronym{osas}{OSAS}{orthonormal subspace alignment scheduling}
\newacronym{nr}{NR}{New Radio}
\newacronym{papr}{PAPR}{peak-to-average power ratio}
\newacronym{pe}{PE}{processing element}
\newacronym{pdsch}{PDSCH}{Physical Downlink Shared Channel}
\newacronym{pf}{PF}{proportional-fair}
\newacronym{phy}{PHY}{physical layer}
\newacronym{qa}{QA}{quantum annealing}
\newacronym{qc}{QC}{quantum computing}
\newacronym{qos}{QoS}{quality-of-service}
\newacronym{qpu}{QPU}{quantum processing unit}
\newacronym{qubo}{QUBO}{quadratic unconstrained binary optimization}
\newacronym{rb}{RB}{resource block}
\newacronym{rbg}{RBG}{RB group}
\newacronym{re}{RE}{resource element}
\newacronym{relu}{ReLU}{rectified linear unit}
\newacronym{sa}{SA}{simulated annealing}
\newacronym{sd}{SD}{sphere decoder}
\newacronym{sdk}{SDK}{software development kit}
\newacronym{sdma}{SDMA}{space-domain multiple access}
\newacronym{sic}{SIC}{successive interference cancellation}
\newacronym{sinr}{SINR}{post-detection SNR}
\newacronym{slp}{SLP}{symbol-level precoding}
\newacronym{soa}{SoTA}{state-of-the-art}
\newacronym{srs}{SRS}{sounding reference signals}
\newacronym{sus}{SUS}{semi-orthogonal user selection}
\newacronym{tdd}{TDD}{time division duplexing}
\newacronym{ue}{UE}{user equipment device}
\newacronym{ula}{ULA}{uniform linear array}
\newacronym{zf}{ZF}{zero forcing}
\newacronym{snn}{SNN}{spiking neural network}
\newacronym{bpsk}{BPSK}{binary phase-shift keying}
\newacronym{qpsk}{QPSK}{quadrature phase-shift keying}
\newacronym{fft}{FFT}{fast Fourier transform}
\newacronym{ifft}{FFT}{inverse fast Fourier transform}
\newacronym{gpp}{GPP}{general purpose processor}
\newacronym{numa}{NUMA}{non-uniform memory access}
\newacronym{simd}{SIMD}{single instruction, mulitple data}
\newacronym{qpi}{QPI}{Intel QuickPath Interconnect}
\newacronym{iq}{I/Q}{in-phase and quadrature}
\newacronym{avx}{AVX}{Advanced Vector Extensions}
\newacronym{oai}{OAI}{OpenAirInterface}
\newacronym{ul}{UL}{uplink}
\newacronym{dl}{DL}{downlink}
\newacronym{tdp}{TDP}{thermal design power}
\newacronym{bios}{BIOS}{Basic Input/Output System}
\newacronym{fr}{FR}{frequency range}
\newacronym{ran}{RAN}{radio access network}
\newacronym{qam}{QAM}{quadrature amplitude modulation}
\newacronym{crc}{CRC}{cyclical redundancy check}
\newacronym{su}{SU}{single-user}
\newacronym{fpga}{FPGA}{field-programmable gate array}
\newacronym{qp}{QP}{quadratic programming}
\newacronym{scs}{SCS}{subcarrier spacing}

%% file: 1_introduction.tex
\Gls{mu}-\gls{mimo} technology has been central to the evolution towards future wireless networks, driving significant connectivity and throughput gains by multiplexing multiple data streams over the same spectral resources. 
However, \gls{mu}-\gls{mimo} processing, specifically \gls{mimo} detection in the uplink and precoding in the downlink, requires substantially higher computational resources compared to that of single-layer transmissions. Identifying signal processing techniques capable of fully exploiting \gls{mu}-\gls{mimo}'s potential in a computationally- and energy-efficient manner remains a significant open question.

It is well understood that when using conventional computational techniques, \Gls{mimo} detection exhibits a trade-off between reliability and complexity. Near one extreme of this trade-off lies \gls{zf} detection, while near the other extreme is the \gls{ml} solution. Some \gls{mimo} detection methods exhibit better reliability-complexity trade-offs than others. Notable examples of the former include \gls{mpnl} \cite{nikitopoulos_massively_2022,jayawardena_g_multisphere_2020}. Still, it is unclear whether alternative computational paradigms, e.g., \gls{qa} \cite{quantumannealing} or deep learning \cite{deeplearning2023}, can break free of this trade-off and offer higher performance than conventional computation methods.


Among emerging paradigms poised to redefine computational efficiency, neuromorphic computing \cite{opportunities_neuro} stands out due to its strong potential for power savings \cite{opportunities_neuro,loihi,truenorth}. Drawing inspiration from the human brain's functionality and energy efficiency, this computing model leverages networks of interconnected computing nodes (neurons), that communicate via event-driven interactions, namely spikes.
While the discussion about \gls{snn} and neuromorphic systems predominantly revolve around their applications in machine learning \cite{opportunities_neuro}, there is an increasing interest in their applicability in non-cognitive problems, with works discussing their applicability on optimization problems such as \gls{qubo}, \gls{qp}  and others \cite{review_noncogn_neur}.
Still, whether it is feasible or efficient to map practical \gls{mimo} \gls{phy} algorithms to neuromorphic, spiking-based architectures remains an open question.
Specifically, critical open questions include whether such approaches can meet the very low \gls{ber} targets demanded by modern mobile communication systems and whether \gls{snn} architectures can adhere to the stringent \gls{phy} latency constraints. 
\setlength{\parskip}{0pt}

In this work, we demonstrate, for the first time, the application of neuromorphic computing to the \gls{mimo} detection problem and discuss the feasibility of ultra-power efficient spiking-based processing in achieving the practical bit error rate targets and latency requirements.
Our evaluations show that neuromorphic processing can achieve slightly better performance than traditional linear detection approaches (i.e., \gls{zf} and \gls{mmse}) 
when the number of base station antennas is more than 4 times the number of MU-MIMO layers. Moreover, for systems where the number of layers approaches the number of base station antennas, the error-rate performance can also match that of conventional linear detectors. This can be achieved by introducing a stochastic input to the \gls{snn}, allowing for a broader exploration of the \gls{mimo} solution space over multiple parallel executions.
Furthermore, the proposed approach can offer substantial reductions of over 45\% in the number of required operations by circumventing entirely the challenge of computationally demanding channel inversions. Moreover, when realized on neuromorphic hardware systems like Intel's Loihi \cite{loihi}
and IBM's True North \cite{truenorth}, the power gains can exceed one order of magnitude compared to conventional processor realizations.
Finally, this work discusses the challenges and the future research directions of \glspl{snn} that could potentially revolutionize PHY processing in next-generation mobile networks.

%% file: 2_system_model.tex
Consider a \gls{mu}-\gls{mimo} uplink transmission model, where a set of $K$ \glspl{ue} transmit data using the same time and frequency resources to a \gls{bs} with $M$ antennas. The complex-valued baseband model is given by
\begin{equation}
    \overline{\mathbf{y}} = \overline{\mathbf{H}} \overline{\mathbf{x}} + \overline{\mathbf{z}} \,,
    \label{eq_mimo_cplx}
\end{equation}
where $\overline{\mathbf{y}}$ is the $M$-dimensional vector of received signals\footnote{In this work, the $\overline{\mathbf{a}}$ notation indicates complex-valued entities, while those without an overline represent real-valued entities.}, $\overline{\mathbf{H}}$ is the $M\times K$ channel matrix, where its elements $\overline{h}_{m,k}$, with $m\in [1,M]$ and $k\in [1,K]$, denote the propagation coefficient from the $k$-th \gls{ue} to the $m$-th \gls{bs} antenna. A flat block channel fading model is used, and it is assumed that the \gls{bs} has perfect knowledge of the channel matrix $\overline{\mathbf{H}}$.
Furthermore, $\overline{\mathbf{x}}$ is the $K$-dimensional vector of transmitted symbols, with each element
taken from the set of an $L$-QAM constellation,
with $E[\overline{x}_k^2]=1$. 
Additionally, $\overline{\mathbf{z}}$ is the $M$-dimensional vector of AWGN noise, with $\overline{\mathbf{z}} \sim \mathcal{CN}\left({\mathbf{0},\sigma_z^2 \mathbf{I}_M} \right)$, where $\sigma_z^2$ is the noise variance. 

The \gls{mimo} model in \eqref{eq_mimo_cplx} can be equivalently represented in standard real-valued form as
\begin{equation}
    {\mathbf{y}} = {\mathbf{H}} {\mathbf{x}} + {\mathbf{z}} \,,
    \label{eq_mimo_real}
\end{equation}
with
\begin{equation}
    \underbrace{\begin{bmatrix}
    \Re(\overline{\mathbf{y}}) \\
    \Im(\overline{\mathbf{y}})
\end{bmatrix}}_{\mathbf{y}} = 
\underbrace{\begin{bmatrix}
    \Re(\overline{\mathbf{H}}) & -\Im(\overline{\mathbf{H}}) \\
    \Im(\overline{\mathbf{H}}) & \Re(\overline{\mathbf{H}})
\end{bmatrix}}_{\mathbf{H}}
\underbrace{\begin{bmatrix} 
    \Re(\overline{\mathbf{x}}) \\
    \Im(\overline{\mathbf{x}})
\end{bmatrix}}_{\mathbf{x}} +
\underbrace{\begin{bmatrix} 
    \Re(\overline{\mathbf{z}}) \\
    \Im(\overline{\mathbf{z}})
\end{bmatrix}}_{\mathbf{z}}
\end{equation}
where $\Re(\cdot)$ and $\Im(\cdot)$ denote the real and imaginary parts, respectively.

The objective of the coherent \gls{mimo} uplink signal detection process is to find a reliable estimate $\widehat{{\mathbf{x}}}$ of the transmitted signals ${\mathbf{x}}$, given that the \gls{bs} receives ${\mathbf{y}}$ and has perfect or estimated knowledge of ${\mathbf{H}}$.

The optimal approach for \gls{mimo} detection when the distribution of the transmitted symbols is uniform, is the \gls{ml} solution
\begin{equation}
    \widehat{{\mathbf{x}}} = \underset{{\mathbf{x}}\in {\mathcal{X}}^{2K}}{\mathrm{arg\, min}} \, \Vert {\mathbf{y}} - {\mathbf{H}} {\mathbf{x}} \Vert^2 \,,
    \label{eq_ml}
\end{equation}
where $\mathcal{X}$ is the set of the PAM symbols corresponding to the real part of the $L$-QAM constellation, with $\vert \mathcal{X} \vert = \sqrt{L}$.
However, \eqref{eq_ml} exhibits complexity that is proportional to $L^{K}$, which can be prohibitively complex when $K$ is large. 

Alternatively, linear detection methods include the \gls{zf} and \gls{mmse} approaches, which are defined as $\widehat{{\mathbf{x}}}_{ZF} = \left( {\mathbf{H}}^T {\mathbf{H}} \right)^{-1} {\mathbf{H}}^T \mathbf{y}$, and $\widehat{{\mathbf{x}}}_{MMSE} = \left( {\mathbf{H}}^T {\mathbf{H}} + \nicefrac{\sigma_z^2}{2} \mathbf{I}_K \right)^{-1} {\mathbf{H}}^T$
respectively, where $(\cdot)^T$ denotes the transpose operator. Although \gls{zf} and \gls{mmse} typically exhibit much lower complexity than the \gls{ml} solution, they offer poor reliability when the channel matrix ${\mathbf{H}}$ is poorly conditioned.

The \gls{ml} and linear detection methods can be seen as extremes in the reliability-complexity tradeoff that is usually exhibited by \gls{mimo} detection methods.

%% file: 3_adopted_methodology.tex
\Glspl{bqm}  are a flexible framework for problems that can be represented in quadratic form, with two primary types: Ising and \gls{qubo}. The main difference is that \gls{qubo} uses binary variables or bits $\mathbf{b}\in\{0,1\}^N $, while Ising employs bipolar variables or spins $\mathbf{s}\in\{-1,1\}^N $. \Glspl{bqm} map these problems onto graphs, with nodes representing variables and edges indicating interactions.

\Glspl{bqm} can be solved through various approaches, such as \gls{qa}, neuromorphic computing, or simulated annealing, each offering unique benefits. 

The Ising formulation for \gls{mimo} detection is described in \cite{quantumannealing}. The Ising and \gls{qubo} formulation are also presented in \cite{kim2019leveraging}. 
However, in the following, we present a simplified \gls{qubo} formulation for the case when the modulation of the transmitted signals $\overline{\mathbf{x}}$ is QPSK. 
Neuromorphic computing makes use of the \gls{qubo} optimization, which is given by 
\begin{equation}
    \widehat{{\mathbf{b}}} = \underset{{\mathbf{b}}\in \{0,1\}^{N}}{\mathrm{arg\, min}} \, \mathbf{b}^T\mathbf{Qb} \,,
    \label{eq_qubo}
\end{equation}
where $N=2K$ is the problem size.

The \gls{ml} objective function of \eqref{eq_ml} can be written as
\begin{align}
    \Vert {\mathbf{y}} - {\mathbf{H}} {\mathbf{x}} \Vert^2 &= \left\Vert {\mathbf{y}} - {\mathbf{H}} \frac{\left(2 \mathbf{b} - \mathbf{1}_{2K} \right)} {\alpha}  \right\Vert^2  \label{eq_qubo_form1} \\
    &= \left\Vert \left( \mathbf{y} + \nicefrac{1}{\alpha} \mathbf{H} \, \mathbf{1}_{2K} \right) - \nicefrac{2}{\alpha} \mathbf{Hb}  \right\Vert^2 \label{eq_qubo_form2}\\
    \begin{split}
        &= \left( \mathbf{y} + \nicefrac{1}{\alpha} \mathbf{H} \, \mathbf{1}_{2K} \right)^T \left( \mathbf{y} + \nicefrac{1}{\alpha} \mathbf{H} \, \mathbf{1}_{2K} \right) \\
        &\phantom{={}} + \nicefrac{4}{\alpha^2} \mathbf{b}^T \mathbf{H}^T \mathbf{Hb} \\
        &\phantom{={}} - \nicefrac{4}{\alpha} \left( \mathbf{y} + \nicefrac{1}{\alpha} \mathbf{H} \, \mathbf{1}_{2K} \right)^T \mathbf{Hb}\,, \label{eq_qubo_form3}
    \end{split}
\end{align}

where the step in \eqref{eq_qubo_form1} converts $\mathbf{x}\in \{\nicefrac{-1}{\alpha}\,\nicefrac{1}{\alpha}\}^{2K}$ to bits $\mathbf{b}\in\{0,1\}^{2K}$, where $\alpha$ is a normalization factor 
which is typically $\sqrt{2}$ for QPSK when 
$E[\overline{x}_k^2]=1$,
and $\mathbf{1}_{2K}$ is the $2K$-dimensional vector of all ones.

Comparing \eqref{eq_qubo} to \eqref{eq_qubo_form3}, and noticing that the first summand in \eqref{eq_qubo_form3} does not depend on $\mathbf{b}$, and therefore can be ignored for optimization purposes, and since $b_j^2 =b_j$, where $j\in [1,2K]$, the \gls{qubo} matrix $\mathbf{Q}$ in \eqref{eq_qubo} is given by
\begin{equation}\label{eq_qubo_matrix}
    \mathbf{Q} = \nicefrac{4}{\alpha^2}  \mathbf{H}^T \mathbf{H} - \mathrm{diag}\left(  \nicefrac{4}{\alpha} \left( \mathbf{y} + \nicefrac{1}{\alpha} \mathbf{H} \, \mathbf{1}_{2K} \right)^T \mathbf{H} \right)\,,
\end{equation}
where the $\mathrm{diag(\mathbf{a})}$ operator outputs a diagonal matrix with the vector $\mathbf{a}$ as its diagonal.

%% file: 4_mapping_to_snn.tex


\floatstyle{spaceruled}
\restylefloat{algorithm}
\begin{algorithm}[t]
\caption{Spiking Network Simulation}
\label{alg_snn}
\small
\begin{algorithmic}[1]
\REQUIRE $T$: Total simulation time, $\Delta T$: Time step size, $R$: Membrane resistance, $U_{\text{th}}$: Firing threshold, $U_{\text{rst}}$: Reset potential
\ENSURE Spikes over time for each neuron $i$ ($S_{t,i})$

\FOR{each time step $t$ $\in$ $[1, T]$}
    \FOR{each neuron $i$ $\in$ $[1, N]$}
        \FOR{each synapse $j$ in $[1, N]$}
            \IF{spike detected at synapse $j$ ($\mathcal{S}_{t,j}=1$)}
                \STATE $I_i \leftarrow I_i + Q_{i,j}$ 
            \ENDIF
        \ENDFOR
        \STATE $U_i(t+\Delta T) \leftarrow U_i(t) + \frac{\Delta t}{\tau} \left( -U_i(t) + R \cdot I_i(t) \right)$
        \IF{$U_i(t) \geq U_{\text{th}}$}
            \STATE $\mathcal{S}_{t,i} = 1$ // \textit{Emit a spike}
            \STATE $U_i(t+\Delta T) \leftarrow U_{\text{rst}}$\\
        \ENDIF
    \ENDFOR
\ENDFOR

\end{algorithmic}
\end{algorithm}

Neuromorphic computing \cite{opportunities_neuro} represents a significant departure from traditional von Neumann architectures, drawing inspiration from the operational principles and structural complexities of biological brains. This computing paradigm integrates processing and memory functions within neurons and synapses, facilitating direct communication through discrete events called spikes. Such an architecture promises substantial improvements in power efficiency, primarily attributed to its massive parallelization capabilities and the event-driven nature of computation \cite{opportunities_neuro,truenorth1}. 
In this section, we present the \gls{snn} architecture we employed for the MU-MIMO detection problem utilizing standard \gls{lif} neurons, explaining its fundamental operations.

\subsection{The Leaky Fire and Integrate Model}
A \gls{lif} neuron model describes the synaptic current dynamics as a linear filter process \cite{brunel_lapicques_2007} that instantly activates when the
membrane potential $U_i(t)$ of neuron $i$ crosses a threshold $U_{th}$. The membrane potential is described by \eqref{eq_neur_diff}, where  $I_i$ is the current injected to the neuron either via electrical stimulation or from other neurons. The time constant $\tau$ is defined as $\tau = RC$ with R being the corresponding resistance and C the capacitance. 
Applying the forward Euler method, \eqref{eq_neur_diff}
can be discretized as presented in \eqref{eq_neur_diff_discr}:
\setlength{\parskip}{0pt}
\begin{equation}
    \tau \frac{{dU_i(t)}}{{dt}} = -U_i(t) + R \cdot I_i(t)
    \label{eq_neur_diff}
\end{equation}
\setlength{\parskip}{0pt}
\begin{equation}
U_i(t + \Delta t) = U_i(t) + \frac{\Delta t}{\tau} (R \cdot I_i(t)-U_i(t))\,.
    \label{eq_neur_diff_discr}
\end{equation}
The injected current $I$ (or synaptic current) for each neuron is evaluated by the summation of the synaptic weights $Q_{i,j}$ of only among synapses that carry a spike towards that neuron. 
The synaptic current $I$ of neuron $i$ at time $t+\Delta t$ is presented in \eqref{eq_neur_current}, where $A_{i,j}\in \{0,1\}$ represents the axon activity of the neuron with value 1 if neuron $i$ received a spike from neuron $j$ and 0 otherwise.
\begin{equation}
    I_i(t+\Delta t) = I(t) + \sum_{j}A_{i,j}\cdot{Q_{i,j}}
    \label{eq_neur_current}
\end{equation}
The \gls{lif} neurons, at each time step, compare the membrane potential with the threshold value $U_{th}$. If the membrane potential is greater or equal to the threshold value, the neuron fires, emitting a spike towards all connected neurons, and resets its membrane potential to a $U_{rst}$ value. 


\subsection{Spiking Network Simulation}
\begin{figure}[t]
    \centering
    { 
        \includegraphics[width=.95\columnwidth]{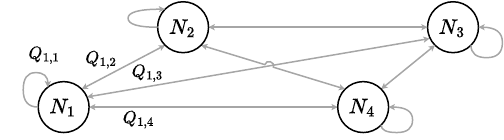}
    }
    \caption{Spiking network connectivity graph for 4 Neurons, with the synaptic weights derived directly from the \gls{qubo} matrix.}
    \label{fig:network_connectivity}
\end{figure}

Algorithm \ref{alg_snn} outlines the basic \gls{snn} functionality predicated on the employed spiking network comprising $N$ interconnected \gls{lif} neurons. As we describe in the following paragraphs, such a network configuration can be employed to derive a potential minima solution for the corresponding MIMO detection problem formulated as a \gls{qubo} optimization problem. The network connectivity and the synaptic weight values are derived from the weighted adjacency matrix $\mathbf{Q}$ in \eqref{eq_qubo_matrix}. Figure \ref{fig:network_connectivity} shows an indicative example with the \gls{snn} connectivity of a network comprised of 4 neurons.
Note that the presented network evolves towards the maximization of the corresponding energy function; therefore, the synaptic weights map to the negated \gls{qubo} matrix.

\begin{figure}[t]
    \centering
    \includegraphics[height=4cm]{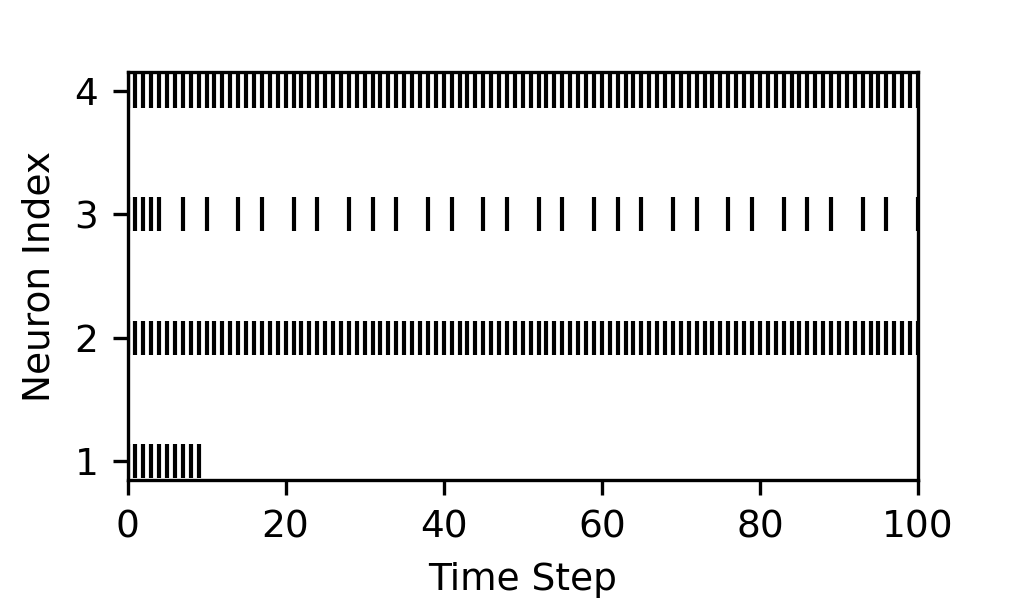}
    \caption{Indicative solution extraction based on the spiking rate of each neuron. The corresponding solution decoded is $\widehat{\textbf{b}}=[0,1,0,1]^T$. Neurons that were active more than 50 percent of the time are decoded as one, while the others are zero.}
    \label{fig:spike_activity}
    \vspace{-10pt}
\end{figure}

Starting from a high input current value $I_i^0 \ \forall i\in[1,N]$, we trigger all the neurons to spike during the initial processing steps. As the processing progresses, the neural dynamics evolve intrinsically. The influence of inhibitory synapses leads to reductions in the membrane potential, subsequently resulting in certain neurons either reducing their spiking frequency or halting their activity in its entirety.
At the end of the processing, after a predetermined number of iterations $T$, the spiking activity of each neuron is linearly translated to the probability that the $i$-th bit of the solution vector $\widehat{\mathbf{b}}$ for the QUBO problem is 1. Figure \ref{fig:spike_activity} shows an example of the spiking activity for an indicative case of 4 interconnected neurons.

%% file: 5_results_and_discussion.tex
\begin{figure*}[ht]
    \centering
    \begin{subfigure}{.32\textwidth}
        \centering
        \includegraphics[height=4cm]{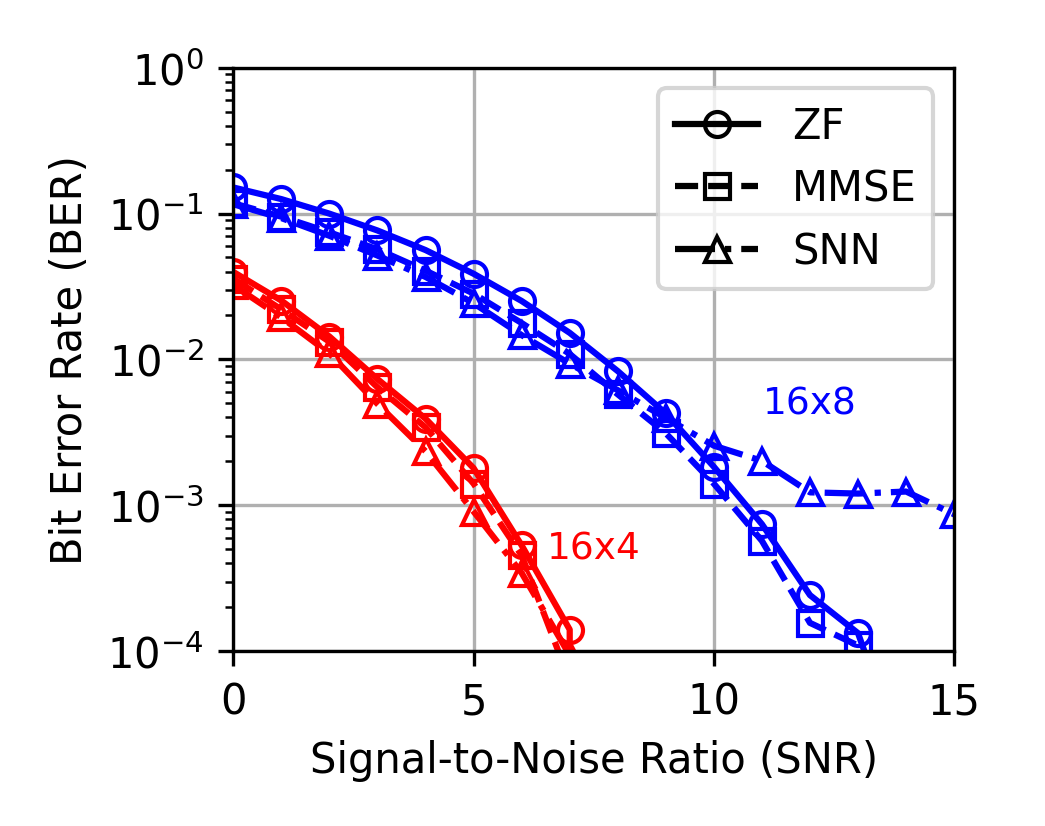}
        \caption{16 Base Station Antennas}
        \label{fig:gls32}
    \end{subfigure}%
    \hfill
    \begin{subfigure}{.32\textwidth}
        \centering
        \includegraphics[height=4cm]{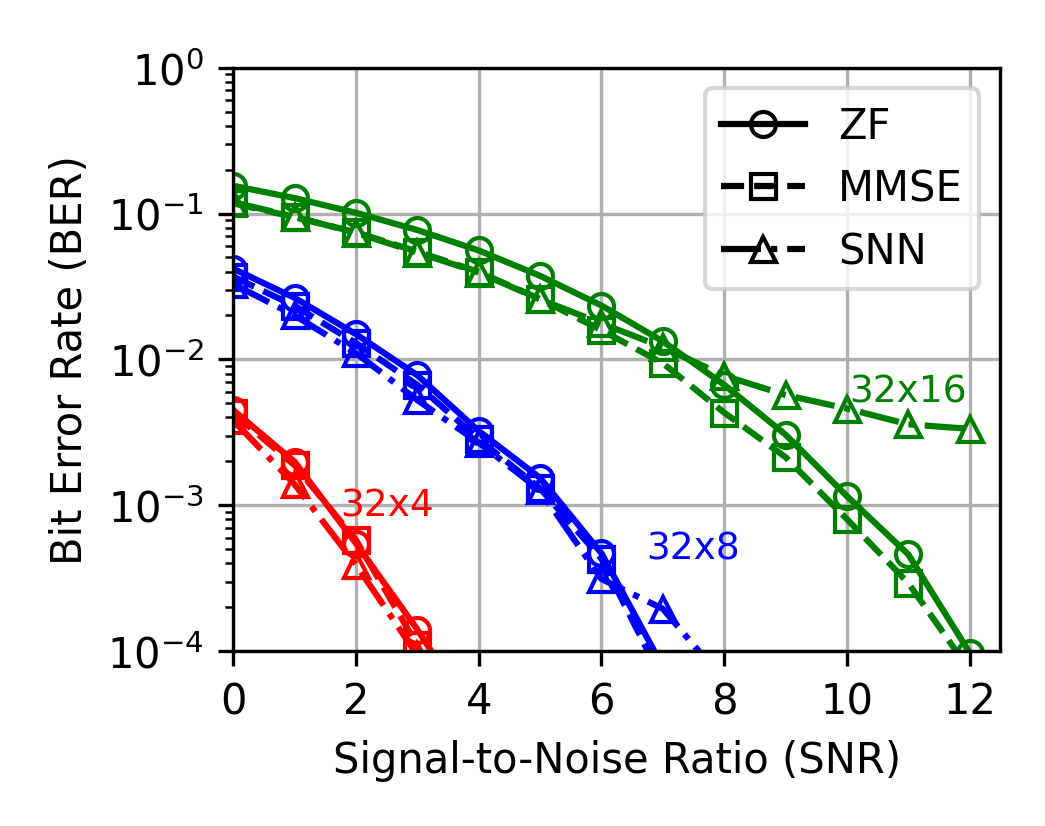}
        \caption{32 Base Station Antennas}
        \label{fig:gls64}
    \end{subfigure}%
    \hfill
    \begin{subfigure}{.32\textwidth}
        \centering
        \includegraphics[height=4cm]{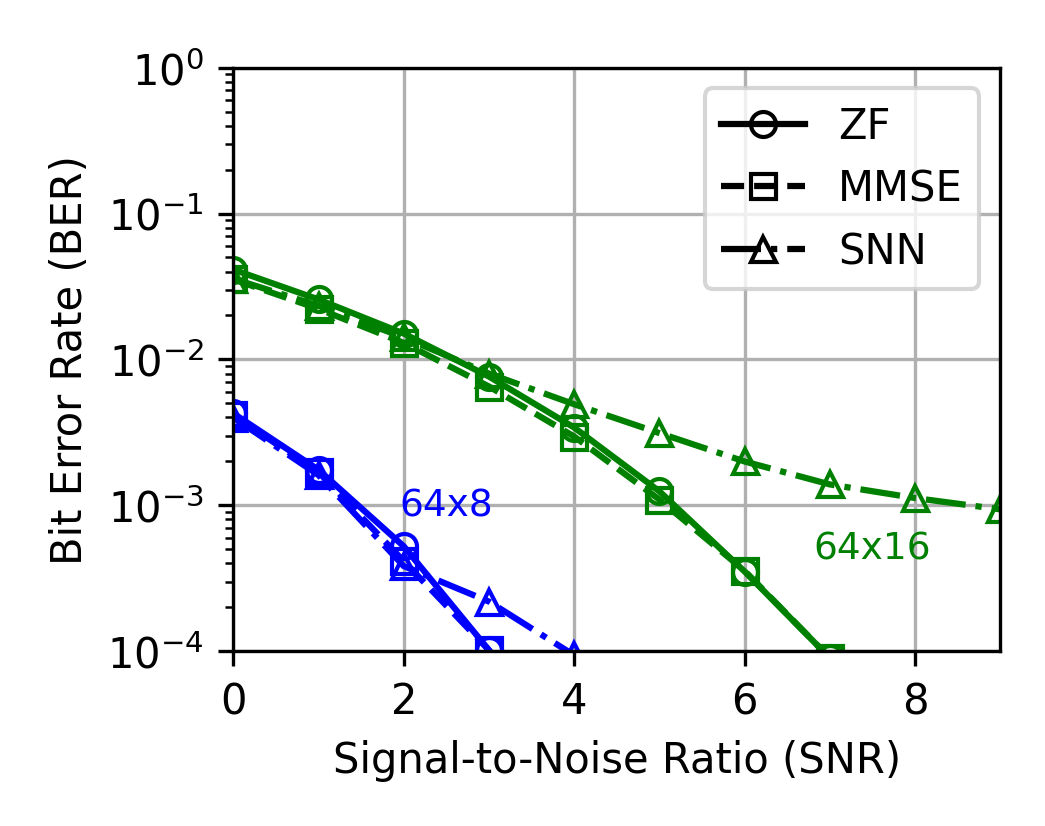}
        \caption{64 Base Station Antennas}
        \label{fig:gls128}
    \end{subfigure}
    \caption{The uncoded \gls{ber} performance of the SNN detector compared to MMSE and Zero Forcing for (a) 16 (b) 32 (c) 64 base station antennas and 4,8 and 16 concurrently transmitted MIMO streams utilizing QPSK modulation.}
    \label{fig:antenna-configs}
\end{figure*}

\begin{figure}
    \centering
    \includegraphics[height=3.5cm]{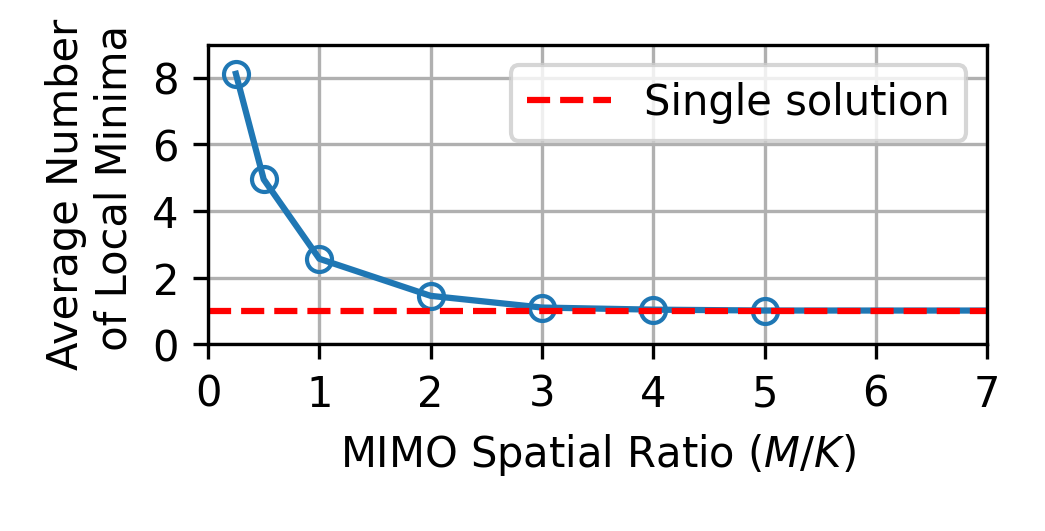}
    \caption{The average number of local minima in the \gls{qubo} energy landscape with the MIMO dimensions}
    \label{fig:locmin}
    \vspace{-10pt}

\end{figure}



\subsection{BER Performance Evaluation}\label{ssec_ber_perf}
To evaluate the SNN's detector error-rate performance, we generated a series of block flat-fading i.i.d. Rayleigh channels corresponding to different \gls{mimo} dimensions. For each setup, we simulated 80 frames of 100 transmissions each. We examined MU-MIMO configurations with 16, 32, and 64 antennas and 4, 8, and 16 MIMO streams within the SNR range of 0 to 15 dB.
For each setup, we evaluated the uncoded \gls{ber} performance of the proposed approach compared to a conventional \gls{mmse} and \gls{zf} detector. The examined system employs QPSK modulation. The spiking network simulation was implemented in Python and the simulation time was set to $T=200$ steps.

The BER results are presented in Figure \ref{fig:antenna-configs}. As shown, in the examined configurations where the spatial ratio (i.e., $M/K$) is higher than 4, the SNN detector slightly outperforms \gls{mmse} reaching the practical \gls{ber} target of $10^{-4}$.
The presence of error floors is directly related to the increase in the number of local minima in the energy landscape when the MIMO spatial ratio ($M/K$) decreases, 
that is, when the number of supported MIMO streams approaches the number of \gls{bs} antennas.
To demonstrate this, Figure \ref{fig:locmin} shows the average number of local minima solutions for different simulated spatial ratios in a system supporting 4 MIMO streams. As shown, the number of local minima in the QUBO energy landscape converges to one with the increase of the spatial ratio, making \gls{snn} detector's performance approach the ML performance in massive MIMO scenarios with many more antennas than spatial streams. 
As we discuss later in Section \ref{ssec_esc_loc_min}, introducing a stochastic input to the system and evaluating multiple generated solutions can lower the corresponding error floor below the BER targets, allowing even smaller spatial ratios to be supported.

\begin{figure}
    \centering
    \includegraphics[height=3.5 cm]{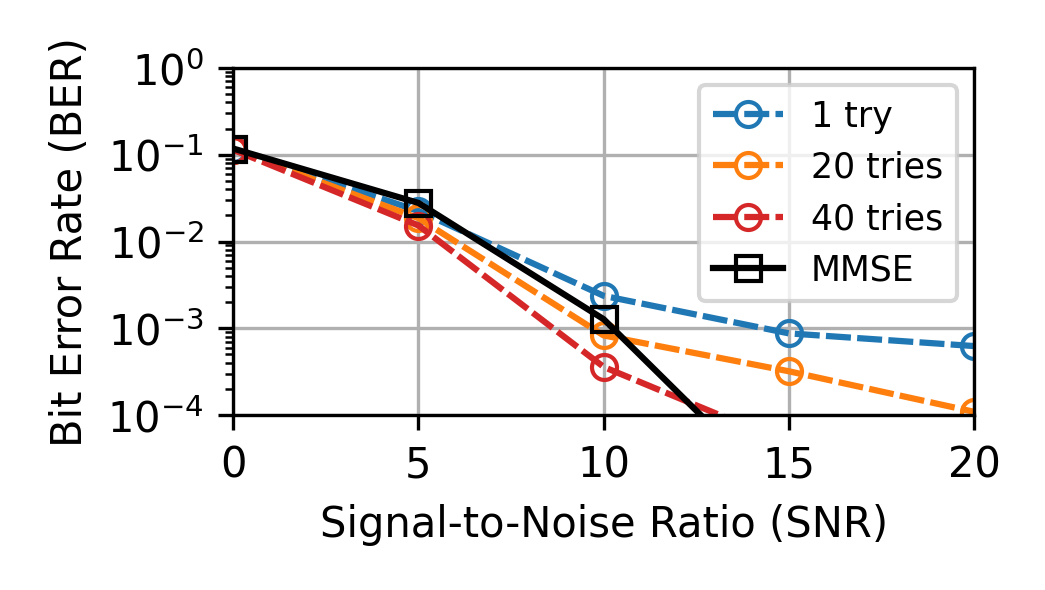}
    \caption{Uncoded BER of SNN detector when multiple iterations and metric evaluations are conducted compared to MMSE for 16x8 MU-MIMO}
    \label{fig:tries_stoch}
    \vspace{-20pt}
\end{figure}

\subsection{Computational Aspects}\label{sec_compl}
Part of the computations, specifically those related to the \gls{qubo} formulation, has to be implemented in traditional processing platforms (e.g., CPU, FPGA). Still, the computational effort required compared to conventional approaches requiring matrix inversion is substantially reduced. Specifically, we find an average reduction of more than 45\% in the total number of operations required across the examined MIMO dimensions. For the calculation we assumed that all operations (i.e., multiplication, addition, division, and square root) are equally weighted (Table \ref{table_operations_qubo}). Those gains render SNN detection computationally favorable, particularly in scenarios where frequent channel inversions are required, e.g., in high mobility cases.
Moreover, it is worth noting that during data transmission, it is not required to update the entire QUBO matrix since the received observable only affects the main diagonal of the QUBO matrix in \eqref{eq_qubo_matrix}, further reducing the overhead of the post-processing.

For the spiking network component of the detector, the employed topology supports highly parallel operation with latency determined by the number of iterations required to extract the solution. In our system, the employed 200 iterations can translate to sub-microsecond latency, with an estimate of approximately 300ns on a pipelined FPGA realization (667MHz), enabling practical real-time application.
Furthermore, we discovered that although error-rate performance can be enhanced with prolonged execution times, it is more efficient from both latency and error-rate perspectives to conduct multiple parallel executions of the same network configuration with a stochastic input in the synaptic current. As detailed in Section \ref{ssec_esc_loc_min}, this approach broadens the search in the energy landscape without significantly affecting the processing latency. This method necessitates the evaluation of the objective function at the end of each run; however, due to the binary nature of the problem, this evaluation can be efficiently performed using only additions in a fully pipelined manner.



Finally, in terms of power consumption, existing neuromorphic platforms such as Loihi \cite{loihi} and TrueNorth \cite{truenorth} have showcased power consumption in the order of milliwatts, making them at least an order of magnitude less power-consuming than modern FPGAs and potentially more than 3 orders of magnitude less consuming compared to modern GPUs and CPUs. To give a first estimate for the proposed MIMO detector assuming a realization on the TrueNorth consumes approximately 323 mW of power with 4096 cores of 256 neurons each. This number of neurons is sufficient to support the parallel detection of an entire uplink 5G \gls{nr} slot (14 OFDM symbols) with 8 QPSK-modulated MIMO streams at 100 MHz of transmission bandwidth (273 resource blocks).

\begin{table}[h]
\centering
\caption{Number of operations for MMSE and QUBO formulations ($k=2K$ $m=2M$)}
\label{table_operations_qubo}
\small
\resizebox{\columnwidth}{!}{%
\begin{tabular}{ c c c c c }
\toprule
Formulation & Multiplications & Additions & Square Roots & Divisions \\ 
\toprule
MMSE (Cholesky Decomp.)   & $2mk^2 + \frac{5}{3}k^3$ & $\frac{4}{3}k^3 + (2m-3)k^2$ & $k$ & $ \frac{k(k-1)}{2}$ \\ 
\midrule
QUBO & $mk^2 + mk$ & $(m-1)k^2 + 2mk - m$ & 0 & 0 \\ 
\bottomrule
\end{tabular}
}
\vspace{-10pt}
\end{table}

%% file: 6_challenges_future_work.tex



\subsection{Scaling to Dense Constellations}\label{ssec_dense_const}
Expanding the \gls{qubo} formulation to QAM modulations higher than QPSK can be challenging because the conventional approach results in a rank-deficient equivalent channel matrix, leading to a higher error floor in the BER curves \cite{quantumannealing}. This happens because 16-QAM and higher-order modulations rely on multi-level-valued symbols instead of bipolar-valued ones.

Several approaches, such as regularization \cite{qa_regularization} and split detection \cite{kim2024x}, have been proposed to address this issue, but they have limitations in solving the underlying problem. Further research into innovative methods is needed to efficiently scale the QUBO formulation to denser constellations.


\subsection{Escaping Local Minima}\label{ssec_esc_loc_min}
As noted in Section \ref{s_results}, local minima in the energy landscape of \gls{qubo} instances can lead to error floors when MIMO configurations have small spatial ratios (i.e., less than 4), exceeding the BER thresholds required for reliable communication.
To address this, a strategy that occasionally permits the system to explore higher energy/illegal states could facilitate overcoming these local minima.
Such an approach can be integrated directly into the spiking network architecture presented in Section \ref{sec_snn_mumimo} by introducing a stochastic input to the synaptic current. For example, by adjusting line 6 of Algorithm \ref{alg_snn},
to $I_i \leftarrow I_i + Q_{i,j} + v$, where $v\sim\mathcal{N}\left( 0, \sigma_v^2  \right)$ is a stochastic variable where its variance $\sigma_v^2$ is a configurable parameter.
Implementing this solution entails running multiple instances of the spiking network in parallel. Following their execution, each instance's result should undergo a full objective function evaluation. As shown in Figure \ref{fig:tries_stoch} for 1, 20, and 40 attempts per \gls{qubo} problem, the error floor drops substantially in the case of a 16x8 MIMO system attaining BER performance akin to that achieved by the MMSE. This realization introduces a trade-off between the implementation area (since multiple parallel networks have to be employed) and the error-rate performance. Still, it’s noteworthy that conducting 40 complete metric evaluations corresponds to an exploration of less than 0.06\% of the total energy landscape of the 16x8 problem.
Future development would include a mechanism for excising soft information from the multiple parallel networks similar to the one proposed for the \gls{mpnl} detector \cite{jayawardena_g_multisphere_2020}.

\section*{Acknowledgement}
This work was supported by the “HiPer-RAN” project, a winner of UK’s Department for Science, Innovation and Technology (DSIT) Open Networks Ecosystem Competition.

%% file: main.bbl
\begin{thebibliography}{10}
\providecommand{\url}[1]{#1}
\csname url@samestyle\endcsname
\providecommand{\newblock}{\relax}
\providecommand{\bibinfo}[2]{#2}
\providecommand{\BIBentrySTDinterwordspacing}{\spaceskip=0pt\relax}
\providecommand{\BIBentryALTinterwordstretchfactor}{4}
\providecommand{\BIBentryALTinterwordspacing}{\spaceskip=\fontdimen2\font plus
\BIBentryALTinterwordstretchfactor\fontdimen3\font minus \fontdimen4\font\relax}
\providecommand{\BIBforeignlanguage}[2]{{%
\expandafter\ifx\csname l@#1\endcsname\relax
\typeout{** WARNING: IEEEtran.bst: No hyphenation pattern has been}%
\typeout{** loaded for the language `#1'. Using the pattern for}%
\typeout{** the default language instead.}%
\else
\language=\csname l@#1\endcsname
\fi
#2}}
\providecommand{\BIBdecl}{\relax}
\BIBdecl

\bibitem{nikitopoulos_massively_2022}
K.~Nikitopoulos, ``Massively {Parallel}, {Nonlinear} {Processing} for {6G}: Potential gains and further research challenges,'' \emph{IEEE Commun. Mag.}, vol.~60, no.~1, pp. 81--87, Jan. 2022.

\bibitem{jayawardena_g_multisphere_2020}
C.~Jayawardena and K.~Nikitopoulos, ``G-{MultiSphere}: Generalizing massively parallel detection for non-orthogonal signal transmissions,'' \emph{IEEE Trans. Commun.}, vol.~68, no.~2, pp. 1227--1239, Feb. 2020.

\bibitem{quantumannealing}
J.~C. De~Luna~Ducoing and K.~Nikitopoulos, ``Quantum annealing for next-generation {MU-MIMO} detection: Evaluation and challenges,'' in \emph{Proc. IEEE Int. Commun. Conf. (ICC)}, 2022, pp. 637--642.

\bibitem{deeplearning2023}
J.~C. De~Luna~Ducoing, C.~Jayawardena, and K.~Nikitopoulos, ``An assessment of deep learning versus massively parallel, non-linear methods for highly-efficient {MIMO} detection,'' \emph{IEEE Access}, vol.~11, pp. 97\,493--97\,502, 2023.

\bibitem{opportunities_neuro}
C.~Schuman, S.~Kulkarni, M.~Parsa, J.~Mitchell, P.~Date, and B.~Kay, ``Opportunities for neuromorphic computing algorithms and applications,'' \emph{Nature Computational Science}, vol.~2, pp. 10--19, 01 2022.

\bibitem{loihi}
M.~Davies \emph{et~al.}, ``Loihi: A neuromorphic manycore processor with on-chip learning,'' \emph{IEEE Micro}, vol.~38, no.~1, pp. 82--99, 2018.

\bibitem{truenorth}
F.~Akopyan \emph{et~al.}, ``{TrueNorth}: Design and tool flow of a 65 {mW} 1 million neuron programmable neurosynaptic chip,'' \emph{IEEE Trans. Comput.-Aided Design Integr. Circuits Syst.}, vol.~34, no.~10, pp. 1537--1557, 2015.

\bibitem{review_noncogn_neur}
J.~Aimone \emph{et~al.}, ``A review of non-cognitive applications for neuromorphic computing,'' \emph{Neuromorphic Computing and Engineering}, vol.~2, Sep. 2022.

\bibitem{kim2019leveraging}
M.~Kim, D.~Venturelli, and K.~Jamieson, ``Leveraging quantum annealing for large {MIMO} processing in centralized radio access networks,'' in \emph{Proc. ACM SIGCOMM}, 2019, pp. 241--255.

\bibitem{truenorth1}
A.~S. Cassidy \emph{et~al.}, ``Cognitive computing building block: A versatile and efficient digital neuron model for neurosynaptic cores,'' in \emph{Proc. IEEE Int. Joint Conf. Neural Netw. (IJCNN)}, Aug. 2013, pp. 1--10.

\bibitem{brunel_lapicques_2007}
N.~Brunel and M.~C.~W. van Rossum, ``\BIBforeignlanguage{en}{Lapicque’s 1907 paper: from frogs to integrate-and-fire},'' \emph{\BIBforeignlanguage{en}{Biological Cybernetics}}, vol.~97, no.~5, pp. 337--339, Dec. 2007.

\bibitem{qa_regularization}
A.~K. Singh, K.~Jamieson, P.~L. McMahon, and D.~Venturelli, ``Ising machines’ dynamics and regularization for near-optimal {MIMO} detection,'' \emph{IEEE Trans. Wireless Commun.}, vol.~21, no.~12, pp. 11\,080--11\,094, 2022.

\bibitem{kim2024x}
M.~Kim, A.~K. Singh, D.~Venturelli, J.~Kaewell, and K.~Jamieson, ``{X-ResQ}: Reverse annealing for quantum {MIMO} detection with flexible parallelism,'' \emph{arXiv:2402.18778}, 2024.

\end{thebibliography}
